\documentstyle[11pt,newpasp,twoside,epsf]{article}
\def\edcomment#1{\iffalse\marginpar{\raggedright\sl#1\/}\else\relax\fi}
\reversemarginpar

\begin{document}
\title{On the Origin of Nitrogen in Damped Ly-$\alpha$ Systems}
 \author{R.B.C. Henry}
\affil{University of Oklahoma, Department of Physics \& Astronomy, Norman, OK  73069 USA}
\author{Jason X. Prochaska}
\affil{UCO/Lick Observatory, UC Santa Cruz, Santa Cruz, CA 95064}

\begin{abstract}
Recent measurements of nitrogen and alpha elements in over 20 damped Ly-$\alpha$ systems (DLA) are compared with similar measurements for numerous emission line objects and stars. It is found that the DLA distribution in the N/$\alpha$-$\alpha$/H plane is bimodal, where most sample DLAs fall along the N/$\alpha$ plateau defined at low Z by dwarf irregulars, while a small group possesses N/$\alpha$ values roughly 0.7~dex less than those on the plateau at similar $\alpha$/H values. We demonstrate with chemical evolution models that a top-heavy or truncated IMF can account for the low N in this second group.
\end{abstract}

\section{Introduction}

Damped Lyman-alpha Systems (DLA) provide an effective abundance probe of the early universe. Ranging in redshift from 1.7-4, they allow us to study the elemental composition of the universe back to within roughly one billion years of the Big Bang. Since DLAs are absorption systems with neutral H column densities $> 2 \times 10^{20}$ cm$^{-2}$ seen against light from background QSOs, their optical depths are large enough to offer a good chance for detecting faint metal lines. Converting these line profiles into ion abundances is relatively straightforward and corrections for unseen ions appear to be minimal, so that inferred gas-phase elemental abundances are reasonably dependable. Abundances of numerous elements including C, N, O, Si, S, Fe, Zn, and Cr are routinely studied and measured in DLAs. Extended discussions about abundances in DLAs are available in Prochaska et al. (2001), Prochaska \& Wolfe (2002), Prochaska et al. (2002; hereafter Pr02), Pettini et al. (2002), and Molaro's review article in this conference. 

The use of DLAs to probe the early production of nitrogen is of particular interest. Because temperatures required for its synthesis are relatively low ($\ge$15 million K), nitrogen may be produced within the H-burning cores and shells of stars representing a broad mass range. There is strong empirical (Henry, Kwitter, \& Bates 2000) and theoretical (van~den~Hoek \& Groenewegen 1997) evidence that intermediate mass stars (IMS; 1$\le$M$\le$8~M$_{\odot}$) produce prodigious quantities of nitrogen. In addition, Henry et al. (2000; hereafter HEK) and Chiappini et al. (2003) have analyzed extant data and concluded that IMS are the principal sources of N in the universe. Using the same data Izotov \& Thuan (1998) have argued that massive stars produce the bulk of the nitrogen when metallicity is low, while Pilyugin et al. (2003) have recently suggested that the data do not allow us to say for certain whether IMS or massive stars are the principal N production sites. So we really don't know for sure where N comes from. If Izotov \& Thuan are correct, then N and other metals produced in massive stars are released into the ISM simultaneously. However, if IMS produce the bulk of the N, then its production lags behind most other elements produced in massive stars because of the finite evolutionary lifetimes of IMS, a time which HEK have estimated to be about 250~Myr. This delay could be longer if rotation or low metallicity lower the effective mass range for N production.

All of this means that nitrogen is a particularly interesting element to track over time, since it's evolution with respect to elements from massive stars can help us gauge the true lag time of N production, and establish the stellar source, i.e. IMS or massive stars.
Because DLAs probe early universe chemical evolution, studies of their N abundances can help us gauge N production in systems of low metallicity which are experiencing their first episodes of star formation.

\section{The Data}

Recently, Pettini et al. (2002) and Pr02 have published N abundances for over 20 DLAs, along with abundances for Si and S, which, because of their massive star origins, serve as oxygen surrogates and track total stellar chemical processing of gas. Fig.~1 is a plot of [N/$\alpha$] versus [$\alpha$/H], where the bracketed values refer to logarithmic quantities normalized to their respective solar values (Grevesse et al. 1996). The filled circles (established values) and triangles (upper/lower limits) represent a combined sample from Pr02 and Pettini et al. The letter symbols show H~II region and dwarf irregular galaxy data, while stars show stellar data, all compiled and explained in HEK. Note that O, Si, and S are collectively referred to as alpha elements, since their abundances appear to scale with each other.  

\begin{figure}[h]
\plotfiddle{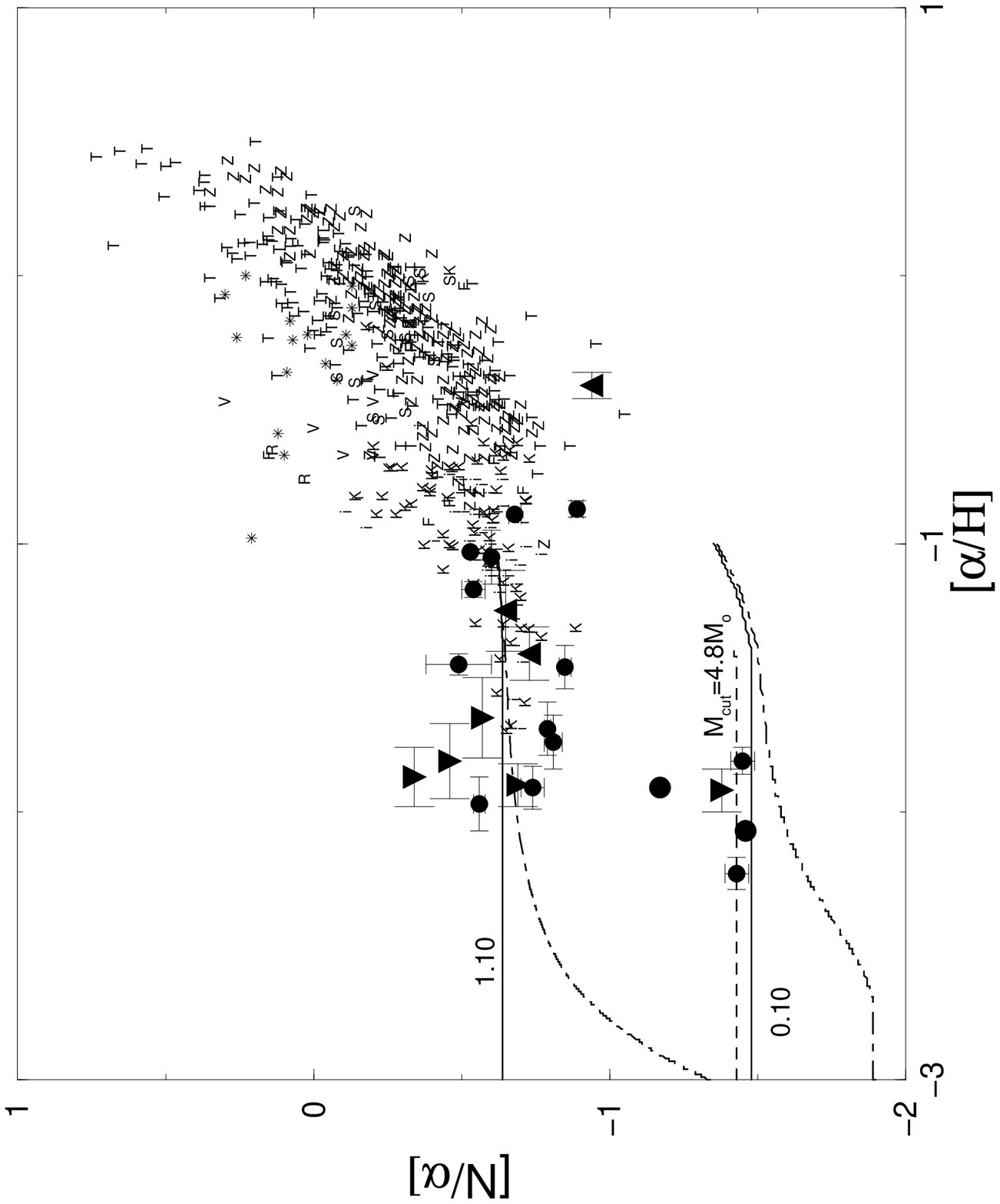}{3in}{270}{50}{50}{-155}{270}
\caption{N/{$\alpha$} versus {$\alpha$}/H expressed logarithmically and normalized to the solar values for DLAs (filled symbols), emission line systems (letters), and stars (star symbols). Model tracks are explained in the text.}
\end{figure}

Fig.~1 can be interpreted as follows. The abcissa gauges the extent of chemical processing of a system, so objects located farther to the right have experienced a greater amount of star formation and alpha element production over their lifetimes. Ordinate values provide a measure of N nucleosynthesis relative to that of the alpha elements. The loose pattern formed by objects other than DLAs suggests that N/$\alpha$ is roughly constant at low metallicity, but increases at higher metallicity values as metallicity climbs. This is consistent with the idea that nitrogen follows a primary-like behavior at low metallicities but a secondary-like behavior at higher values.

Conclusions relating to the DLA distribution are: 1)~Most DLAs fall along the primary plateau at [N/$\alpha$]$\approx$-0.75; but 2)~a small number fall significantly below the plateau, albeit at similar metallicity values as those in the first group, with a centroid value of [N/$\alpha$]$\approx$-1.4.  This apparent bimodality in the DLA distribution cannot be explained by dust or unseen ions of nitrogen.

\section{Explaining the Bimodality}

Clearly a group of five objects such as the low nitrogen DLAs (LN-DLA) represents a statistically small sample, and it is possible that the distinction between the two DLA groups will become more blurred as more N abundances are measured in DLAs. But the empirical result is nevertheless intriguing. So, in what follows we will assume that the bimodal DLA distribution is real and that IMS produce most of the nitrogen in the universe. We now explore four possibilities for explaining the DLA bimodality.

The first explanation is that we are witnessing the effects of time delay in N production (HEK; Pettini et al. 2002). Following initial star formation, massive stars die quickly and expel alpha elements along with a small amount of N into the system's interstellar medium. Then after a delay, primary N from IMS is expelled and N/$\alpha$ rises to the primary plateau. The LN-DLAs, therefore, may be objects in which IMS have not yet released their N, so their N/$\alpha$ levels are determined only by massive star N production.

Fig. 2 shows the temporal evolution of [N/$\alpha$] after a star burst, as predicted by a chemical evolution model similar to those discussed in HEK. Notice the abrupt rise in [N/$\alpha$] to a small plateau value of around -1.8. The plateau arises because massive stars have ceased their N production but IMS have not yet begun to make their contribution to the interstellar medium. However, as IMS begin to expel their N at around 60~Myr, [N/$\alpha$] resumes it upward trend until it reaches the primary plateau value near -0.7. Notice that the rise to the plateau value is gradual rather than abrupt, forecasting a smooth distribution of objects between the LN-DLA location and the primary plateau, contrary to what is observed. On the other hand, a bimodal distribution in [N/$\alpha$] is more consistent with a sudden rise in N production, so that there would be only a small probability of observing DLAs between the primary plateau and the observed region of LN-DLAs. An example of such behavior is shown with a dashed line in Fig.~2. If the gap becomes filled in with additional objects in the future, then the time delay scenario would become feasible. For now, it appears to be inconsistent with a bimodal distribution.

\begin{figure}[h]
\plotfiddle{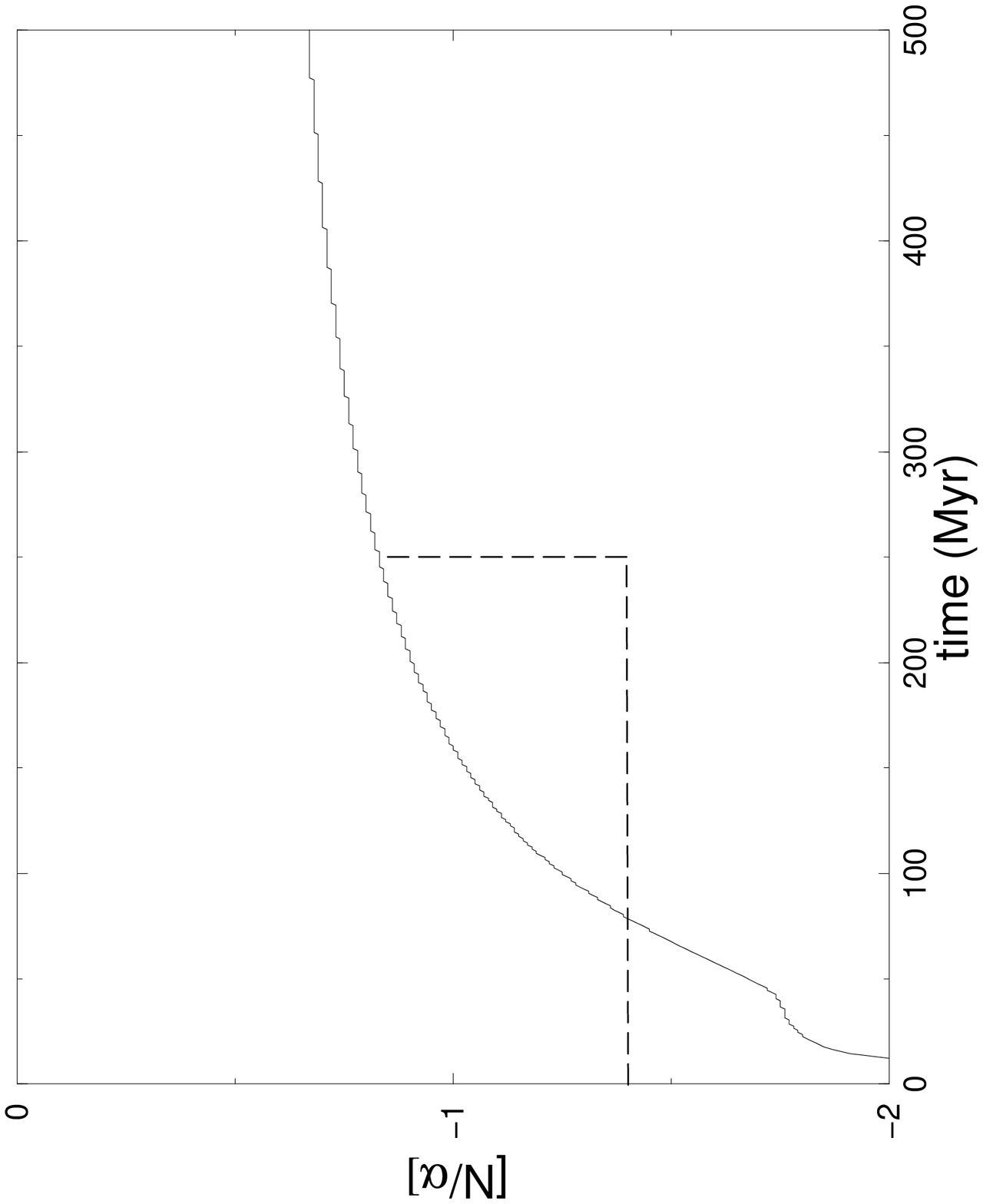}{2in}{270}{35}{35}{-125}{185}
\caption{[N/$\alpha$] versus time for a chemical evolution model similar to those discussed in HEK (solid line). The dashed line shows an example of an abrupt release of nitrogen.}
\end{figure}

A second possibility for explaining the observed DLA distribution relates to the metallicity sensitivity of stellar yields. A simple closed-box model of chemical evolution can be used to show that the effective yield\footnote{The effective yield of element X, $P_X$, is the stellar yield as a function of mass integrated over a Salpeter IMF. See HEK for details.} ratio of N/O needed to explain the observed [N/$\alpha$] of -1.4 of the LN-DLAs is $P_N/P_O = 0.005,$ henceforth called the ``empirical'' value. The left panel of Fig.~3 shows a comparison of this empirical ratio with predictions of two yield sets. The first set combines the yields of van den Hoek \& Groenewegen (1997) for IMS and Woosley \& Weaver (1995) for massive stars (HG+WW), while the second set links Marigo's (2001) IMS yields and Portinari et al.'s (1998) massive star yields (MP). The first set results in a value which is 1.5~dex above the empirical level, while the second set predicts a number which is 0.9~dex above the empirical value. So current ratios of effective yields appear significantly larger than what is required to explain the LN-DLAs. Additionally, it would be difficult to understand how a stellar population characterized by a single metallicity could produce a bimodal distribution in [N/$\alpha$] from a single set of yields unless one introduces another parameter such as rotation into the picture (Langer et al. 1997; Maeder \& Meynet 2001).

\begin{figure}[h]
\plotfiddle{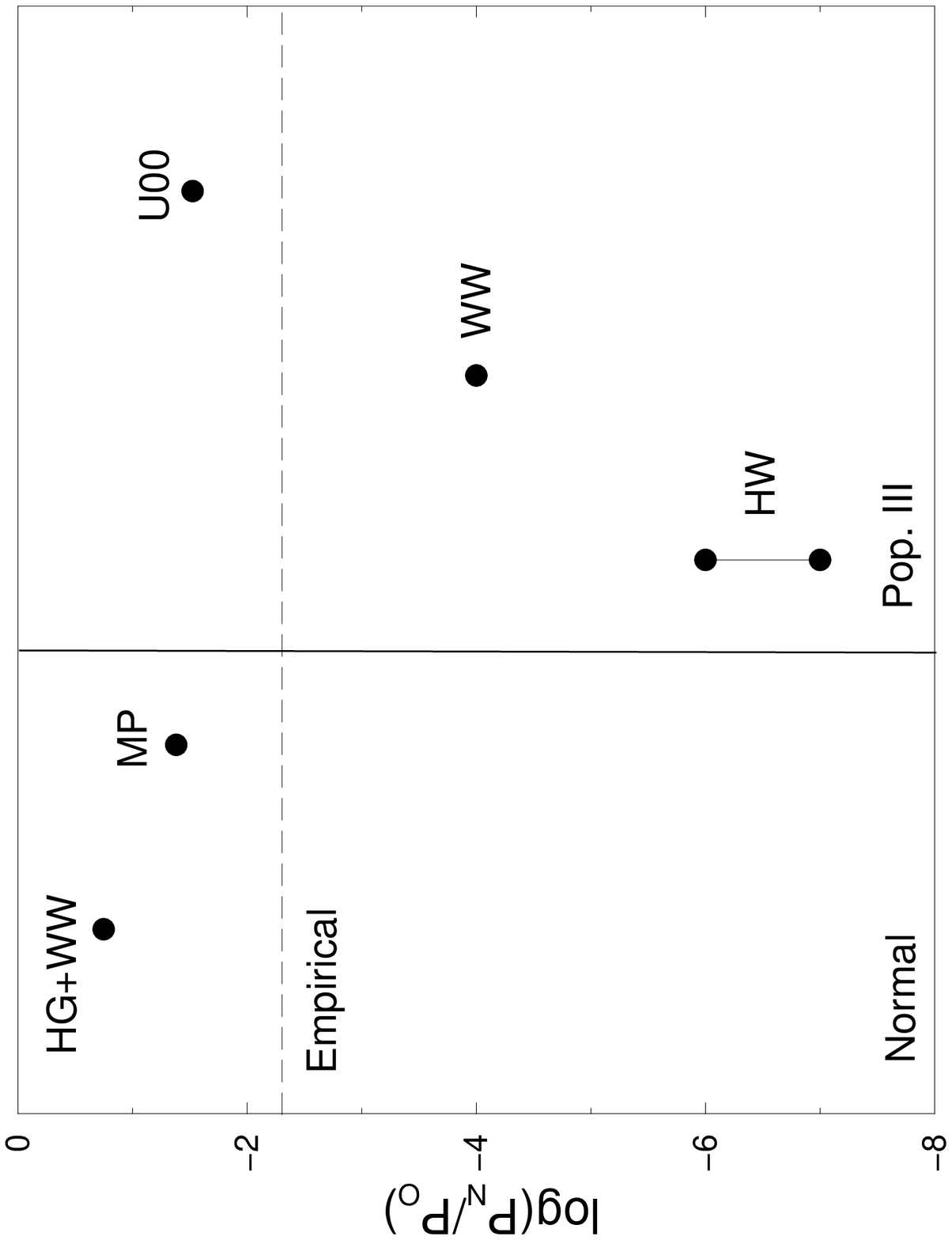}{2in}{270}{35}{35}{-125}{185}
\caption{Effective yield ratios of N/O for a low metallicity normal stellar population (left panel) and a zero metallicity Population~III stellar group (right panel) as determined from yield predictions published by sources discussed in the text. The dashed line shows the empirical value inferred from observations with a simple model of chemical evolution. The horizonatal distribution of the points has no significance.}
\end{figure}

A third explanation for the LN-DLAs may come from nucleosynthesis occurring at zero metallicity, i.e. from Population~III stars. Yields for such star groups have been calculated by Woosley \& Weaver (1995; 13-40~M$_{\odot}$; WW), Umeda et al. (2000; 13-25~M$_{\odot}$; U00), and Heger \& Woosley (2002; 140-260~M$_{\odot}$; HW) for the indicated mass ranges. The results from these yields are plotted in the right panel of Fig.~3. The predictions of Woosley \& Weaver and Heger \& Woosley fall far below the empirical value, while the predictions of Umeda et al. exceed it.\footnote{For HW, we simply show a range in ratio values inferred directly from their yield tables by inspection.} Currently, the Population~III yield predictions are inconsistent with the observed [N/$\alpha$] level, although investigations of these stars are in an early stage. Thus, future work may still prove that this mechanism is feasible.

The fourth hypothesis, and the one that we currently favor, involves altering the IMF. We experimented with both top-heavy (flatter) IMFs and IMFs which are truncated below a certain mass threshold. In both cases this reduces the IMS:massive star ratio and causes the N/O ratio to fall noticeably, since we are assuming that IMS produce the bulk of the N. We calculated simple numerical chemical evolution models of closed systems using the program described in HEK along with the MP yield set and experimented with various IMF slopes and mass thresholds. The solid lines in Fig.~1 correspond to models which assumed instantaneous recycling, i.e. no time lag between IMS and MS evolution, while the dot-dashed lines show the effect of relaxing this assumption. The top curves refer to an IMF close to the Salpeter relation (slope=1.35). Meanwhile, the bottom curves show what happens when the IMF slope is flattened to 0.10. In addition, the dashed line refers to a model in which the IMF is truncated below 4.8M$_{\odot}$. Similar results are obtained if these experiments are repeated using the HG+WW yield set. Clearly, a flat or truncated IMF is consistent with the LN-DLAs. It is interesting to note that the popular view of Population~III stars comprising only massive objects is consistent with the idea of a truncated IMF.

What we are proposing, then, is that star formation at low metallicity produces a relatively top-heavy population of stars in some systems, resulting in a N-deficient enrichment compared to the standard primary ratio. Such a scenario offers a straightforward mechanism for producing the low [N/$\alpha$] value observed for the LN-DLAs. If correct, these new observations may offer the first evidence that the mass spectra of some early stellar populations were much different than what we observe today. 

\acknowledgments

This research was supported by NSF grant AST-9819123

\end{document}